\documentclass[twocolumn,aps,pre,amsmath,amssymb,longbibliography]{revtex4-1}
\usepackage{graphicx}% Include figure files
\usepackage{dcolumn}% Align table columns on decimal point
\usepackage{bm}% bold math
\usepackage{multirow}
\usepackage{amssymb}
\usepackage{amsmath}
\usepackage{graphicx}
\usepackage{xcolor}
\usepackage[colorlinks,citecolor=blue,linkcolor=red,urlcolor=blue]{hyperref}
\usepackage{indentfirst}
\usepackage{amsmath}
\usepackage{cases}

\begin{document}
\title{Driven dynamics of localization phase transition in the Aubry-Andr\'{e} model with initial gapless extended states}
\author{Xin-Yu Wang\textsuperscript{1,2}}
\altaffiliation{These authors contribute equally to this work.}
\author{Wen-Jing Yu\textsuperscript{1,2}}
\altaffiliation{These authors contribute equally to this work.}
\author{Yue-Mei Sun\textsuperscript{1,2}}
\author{Liang-Jun Zhai\textsuperscript{1,2}}\email{zhailiangjun@jsut.edu.cn}
\affiliation{\textsuperscript{1}The School of Mathematics and Physics, Jiangsu University of Technology, Changzhou 213001, China}
\affiliation{\textsuperscript{2}The Jiangsu Key Laboratory of Clean Energy Storage and Conversion, Jiangsu University of Technology, Changzhou 213001, China}
\date{\today}
\begin{abstract}
Recently, the driven dynamics of localization phase transitions have garnered growing interest.
However, studies so far have mainly considered initial localized states, whose driven dynamics follow the Kibble-Zurek mechanism (KZM).
In this study, we investigate the driven dynamics of the localization phase transition in the Aubry-Andr\'e (AA) model starting from a gapless extended state, which violates the adiabatic-impulse scenario of KZM.
By linearly driving the quasiperiodic potential strength across the critical point, we numerically simulate the driven dynamics and analyze the scaling behavior of both the inverse participation ratio ($\mathcal{I}$) and the dynamic deviation from the instantaneous ground state energy $(\mathcal{D})$.
We demonstrate that the driven dynamics starting from initially extended states satisfies the criterion for the applicability of KZM and its extension, finite-time scaling (FTS).
The scaling functions governing the driven dynamics of both $\mathcal{I}$ and $\mathcal{D}$ have been derived based on FTS and numerically validated.
We found that the scaling functions exhibit significant differences at large $R$ and small $R$, and also differ considerably from the scaling functions when the initial state is localized, highlighting the crucial role of initial state behavior.
The established scaling laws remain robust across a wide range of system sizes and driving rates, providing testable predictions for experimental realizations.
\end{abstract}
\maketitle

\section{Introduction}
The localization phenomenon remains a perennial focal point in the study of quantum condensed matter~\cite{Anderson1958,Abrahams1979,Kramer1993,Evers2008,Lee1985,Schwartz2007,Evers2008,Segev2013,Wiersma2013,Alexey2023}.
Foremost among these is disorder-induced Anderson localization, although it fails to predict the localization transition in lower dimensions~\cite{Abrahams1979}.
In recent years, quasiperiodic systems have attracted significant interest, as they host unique manifestations of entirely new phases of matter and exhibit transitions between them~\cite{Aubry1980,Biddle070601,HepengYao2019,Schirmann2024,Ribeiro2024,Jan2023,Agrawal2020,Agrawalprl2020,Goblot2020,Roy2022,Agrawal2022, Zhou2023,Smith176601}.
The Aubry-Andr\'{e} (AA) model is a canonical example of quasiperiodic systems~\cite{Aubry1980}.
Based on the AA model and its extensions, researchers have conducted extensive studies on localization phenomena and related topics~\cite{Kraus2012,Guo2024,Biddle2011, Sahoo2025, Yi2025,Xuan2022,Ting2022,Saha2016,Ganeshan2015,Wangprl2020,Hu2025,Liu2022,Gao2024,Banerjee2025,Iyer2013,Schreiber2015, Mastropietro2015,Khemani2017,Zhang2018,Rispoli2019,Weiner2019,Xu2019,zhai2020,Cookmeyer2020,Strkalj2021,Tang2021,Strkalj2022,Prasad2024, Huang2024, Chakrabarty2025,Zeng2017,Jiang2019,Longhi20192,Longhi224206,Liu2021,zhai2021,Guo2021,xueprl2022,Acharya2024,Guocx2024,Zeng2024,Sun2024,Chen2025}, such as the mobility edges~\cite{Zhou2023,Saha2016,Ganeshan2015,Wangprl2020,Hu2025,Liu2022,Gao2024,Banerjee2025}, the interplay between localization and interaction~\cite{Iyer2013,Schreiber2015,Mastropietro2015,Khemani2017,Zhang2018,Rispoli2019,Weiner2019,Xu2019,zhai2020,Cookmeyer2020,Strkalj2021, Tang2021, Strkalj2022,Prasad2024,Huang2024,Chakrabarty2025}, and the interplay between localization and non-Hermitian physics~\cite{Zeng2017,Jiang2019,Longhi20192,Longhi224206,Liu2021,zhai2021,Guo2021,xueprl2022,Acharya2024,Guocx2024,Zeng2024,Sun2024,Chen2025}.

Meanwhile, rapid experimental progress has stimulated increasing interest in the non-equilibrium dynamics of quantum systems~\cite{Blatt2012,Mukherjee2017,Gross2017,Ebadi2021,xueKZM2021}.
Research in this area not only promotes the exploration of novel states of matter, but also allows for precise manipulation and preparation of quantum states~\cite{Bedow2022,Perfetto2021,Mankowsky2016,Popkov2013,Cheng2023,Giudici2022}.
In particular, the study of driven dynamics has garnered significant attention due to its relevance to adiabatic quantum computation and programmable quantum devices~\cite{Keesling2019,Dupont2022,Reichhardt2022,King2023,Soto2024}.
Within theoretical frameworks addressing driven dynamics, the Kibble-Zurek mechanism (KZM) has been developed to characterize dynamical behavior near critical points~\cite{Kibble1976,Zurek1985,Zurek2005,Dziarmaga2005,Dziarmaga2010,Polkovnikov2011,Yin2017,zhai2018,Shu2025}.
A cornerstone of KZM is the adiabatic-impulse hypothesis, which requires the system to initially reside in a gapped ground state~\cite{Dziarmaga2010,Polkovnikov2011}.
Nevertheless, driven quantum systems with gapless initial conditions have also come under theoretical investigation~\cite{Deng2008,Polkovnikov2008,Pellegrini2008}, prompting the formulation of finite-time scaling (FTS) as a generalization beyond the gapped-state assumption~\cite{Gong2010,Huang2014,Feng2016}.
In FTS, the driving rate serves as a key scaling parameter, leading to a complete scaling description of the full non-equilibrium evolution~\cite{Gong2010}.
Most recently, a universal criterion has been proposed to evaluate the applicability of both KZM and FTS in describing driven dynamics starting from gapless initial states~\cite{Zeng2025}.

There has been growing interest in the non-equilibrium dynamical behavior of localization transitions~\cite{Morales2014,Cheng2399,Bairey2017,Yang2017,Zhou2021,Tiwari2024,Xu2020,Xuzh2021}.
For example, studies have focused on periodically driven dynamics and the characteristics of dynamical localization in such systems~\cite{Morales2014,Bairey2017,Zhou2021,Tiwari2024}.
Furthermore, numerous studies have investigated the driven dynamics of localization transitions~\cite{Sinha2019,Xuan2023,Liang2024,Zhai2022,Zhaifphy,Sun2025}.
It has been found that when the system is slowly driven from an initial state far from the critical point, the dynamics of the localization transition can be well described by the KZM~\cite{Sinha2019,Zhai2022}.
Moreover, in systems where multiple localization mechanisms coexist, e.g., the coexistence of disorder and AA potential, researchers have found that the driven dynamics also obey a hybrid KZM~\cite{Xuan2023,Liang2024,Sun2025}.

In previous studies, researchers have primarily focused on the driven dynamics of localization phase transitions in sufficiently large systems with initially localized states~\cite{Sinha2019,Zhai2022}.
However, theoretical investigations into driven dynamics starting from extended states remain scarce.
Since extended states represent a type of gapless initial state, it is particularly meaningful to investigate whether driven dynamics starting from such states can still be described by KZM and FTS.
Furthermore, comparing the dynamical behavior in these systems with that originating from localized initial states offers valuable insights into the role of initial state topology in non-equilibrium quantum evolution.

In this work, we employ the one-dimensional AA model to study driven dynamics of localization transition starting from a gapless extended state.
We demonstrate that the driven dynamics with initial extended states in this system satisfies the criterion for the applicability of KZM and FTS.
By linearly increasing the strength of the quasiperiodic potential across the localization transition point, we numerically simulate the driven dynamics of the localization transition starting from an extended state.
The inverse participation ratio (${\mathcal I}$) and the dynamic deviation from the instantaneous ground state energy $(\mathcal{D})$ are employed as order parameters, the scaling functions governing the behavior of the ${\mathcal I}$ and $\mathcal{D}$ are proposed based on FTS and numerically validated.
We observe that the scaling functions of these order parameters exhibit distinct forms in the small and large driving rate regimes,
and they differ significantly from those observed in dynamics initiated from localized states.
The scaling laws established in this work remain robust over a wide range of system sizes and driving rates.
Given recent experimental advances in realizing the AA model, the proposed scaling relations are expected to be readily testable in experimental settings.

The rest of the paper is arranged as follows.
In Sec.~\ref{modelphase}, the model of AA model is introduced.
In Sec.~\ref{dynamics}, we present the scaling equations of the driven dynamics of both the ${\mathcal I}$ and ${\mathcal D}$ when starting from extended states, and numerically validate of these equations.
Finally, a brief summary is presented in Sec.~\ref{sum}.

\section{\label{modelphase}The AA Model}

The Hamiltonian of the AA model reads~\cite{Aubry1980}
\begin{eqnarray}
\label{Eq:model}
H=-\sum_{j}^{L}{J( c_{j}^\dagger c_{j+1}+h.c.})+\lambda\sum_{j}^{L}{w(j)c_j^\dagger c_j}.
\end{eqnarray}
Here, $c_j^\dagger (c_j)$ is the creation (annihilation) operator of the hard-core boson on $j$ site, and $L$ is the system size.
$w(j)=\cos{[2\pi(\gamma j+\phi)]}$ with $\gamma$ being an irrational number is the quasi periodic onsite potential,
and $\lambda$ measures the amplitude of the quasi-periodic potential.
$\phi\in[0,1)$ is phase of the potential.
We assume $J=1$ as the unity of energy, and $\gamma=(\sqrt{5}-1)/2$.
To satisfy the periodic boundary condition of the quasi-periodic potential, we approximate $\gamma$ as a rational number $F_n/F_{n+1}$ where $F_{n+1}=L$ and $F_{n}$ are the Fibonacci numbers~\cite{Jiang2019,Zhai2022}.

\begin{figure}[htbp]
\centering
  \includegraphics[width=2.3in,clip]{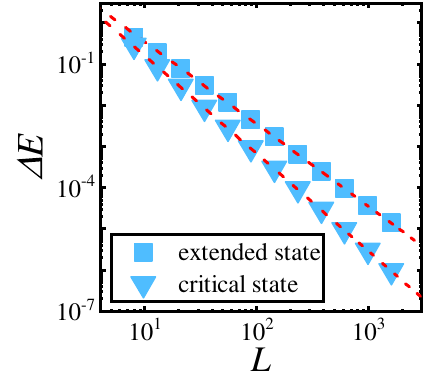}
  \vskip-3mm
  \caption{Energy gap $\Delta E$ as a function of $L$ for the extended and critical states. We use $\lambda=\lambda_c-1$ for the extended state as an example.
  The results are averaged for $1000$ samples of $\phi$. The log-log coordinate is used.}
  \label{EnergyGap}
\end{figure}

The AA model has a localization-extended transition point at $\lambda_c=2$.
When $\lambda<\lambda_c$, all its eigenstates are extended; whereas when $\lambda>\lambda_c$, all eigenstates are localized.
Another important feature of the AA model is that when the system in the extended phase, the energy gap between ground and first excited state ($\Delta E$) becomes zero in the thermodynamic limit of $L\rightarrow \infty$, resulting in a gapless state.
These gapless states are also considered to be a series of continuous critical states~\cite{Zeng2025}.
When finite-size effect is taken into account, $\Delta E$ of the extended state exhibits a power-law dependence on system size, scaling as $L^{-z'}$, where $z'>0$ is the dynamical exponent.
Note that this exponent for the extended state differs from the dynamical exponent of the critical state.

As shown in Fig.~\ref{EnergyGap}, we present $\Delta E$ as a function of $L$ for both the extended and critical states.
It can be observed that the $L$ dependence of $\Delta E$ for both states exhibits linear behavior on a log-log scale.
The exponent obtained from the power-law fitting for the extended state is $z'=2$, while the dynamical exponent for the critical state is $z = 2.37$~\cite{Sinha2019,Wei2019}.

\section{\label{dynamics}the driven dynamics with initial extended states}
\subsection{General theory of driven dynamics with initial gapless states}

In this section, let us  briefly introduce the theoretical framework for driven dynamics starting from a gapless initial state.
We assume that the initial value of $\lambda$ is far from the critical point $\lambda_c$ and that the system is driven linearly across the critical point $\lambda_c$ at a constant rate $R$.
Defining $\varepsilon\equiv \lambda-\lambda_c$ as the distance to the critical point, the driving process satisfies
\begin{equation}\label{driven}
 \varepsilon = \varepsilon_0 + Rt,
\end{equation}
where $\varepsilon_0 = \lambda_0 - \lambda_c$ represents the initial distance to the critical point.

In the conventional quantum version of KZM theory, the driven dynamical process is divided into the adiabatic evolution regime and the impulse regime~\cite{Dziarmaga2010,Polkovnikov2011,Sinha2019}.
For $|\varepsilon|>R^{1/r\nu}$ with $\nu$ being a critical exponent and $r=z+1/\nu$, the evolution is in the adiabatic regime,
and the system evolves adiabatically because the state has sufficient time to adapt to changes in the Hamiltonian.
However, when the system enters the impulse regime of $|\varepsilon|<R^{1/r\nu}$, the state stops evolving due to critical slowing down.
Therefore, the frozen state is no longer the ground state but an excited state.
This scenario can be comprehensively described by the scaling functions of KZM.

For the driven dynamics with the initial state being a gapless state, the situation differs fundamentally.
Due to its gapless nature, the system's evolution deviates from adiabaticity even during the initial stages.
Excitations can be copiously produced in the initial gapless phase and subsequently carried into the critical region, influencing the nonequilibrium properties near the quantum critical point.
Consequently, the conventional Kibble-Zurek approximation of distinct adiabatic and impulse regimes breaks down in this scenario.

However, recent studies on driven dynamics starting from a gapless initial state have revealed that if the system satisfies the following precondition~\cite{Zeng2025}
\begin{equation}\label{criterion}
  z'<r,
\end{equation}
where $z'$ is the dynamical exponent of gapless state,
the driven dynamics can still be described by the KZM and FTS.
This is because, when this condition is met, the dominant nonequilibrium universal behavior originates from the critical region of the quantum critical point.
In this case, the driving rate $R$ with the critical exponent $r$ governs the dynamic scaling behavior.

For the AA model, the dynamical exponent of the extended state is $z' = 2$, while the critical exponents at the quantum critical point are $\nu = 1$ and $z = 2.37$, yielding $r = z + 1/\nu = 3.37$.
This clearly satisfies the aforementioned criterion in Eq.~(\ref{criterion}).
In the following discussion, we take the ground state of the AA model at $\lambda =\lambda_c-1$, which is sufficiently far from the critical point, as the initial state,
and then slowly vary $\lambda$ across the critical point into the localized phase.
The driven dynamic processes are simulated by numerically solving the Schr\"{o}dinger equation for the AA model, and the ${\mathcal I}$ and ${\mathcal D}$ are taken as order parameters to verify the applicability of FTS.

\begin{figure}[htbp]
\centering
  \includegraphics[width=3.5in,clip]{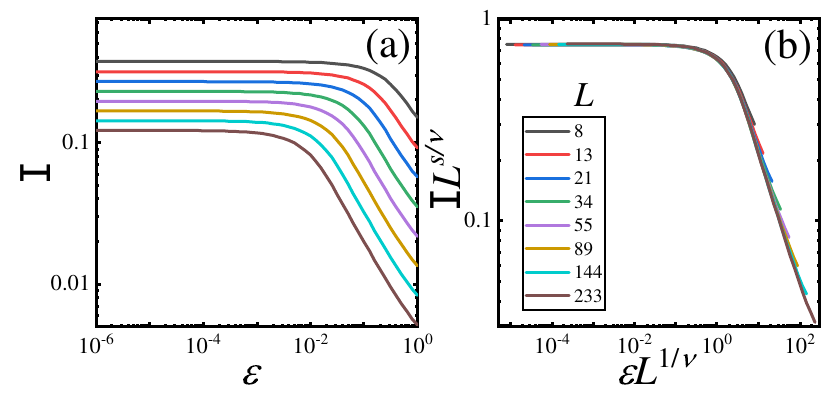}
  \vskip-3mm
  \caption{(a) Static $\mathcal{I}$ versus $\varepsilon$ for different $L$ and (b) the rescaled curves according to Eq.~\eqref{Eq:scaleipr}.
  The results are averaged for $1000$ samples of $\phi$. The log-log coordinate is used.}
  \label{staticIPR}
\end{figure}

\subsection{Driven dynamics of ${\mathcal I}$}
${\mathcal I}$ is defined as~\cite{Bauer1990,Fyodorov1992}
\begin{equation}
\label{Eq:ipr}
{\mathcal I} =\frac { \sum_{j=1}^L|\Psi(j)|^4} {\sum_{j=1}^L|\Psi(j)|^2},
\end{equation}
where $\Psi(j)$ is the wavefunction.
For a localized state, the wave function is localized on some isolated sites, and ${\mathcal I}\propto{L^0}$, whereas ${\mathcal I}\propto L^{-1} $ for the extended states.
For the critical state, it scales as ${\mathcal I} \propto L^{-s/\nu}$ with $s = 0.333$~\cite{Xuan2022}.

Near the critical point, $\mathcal{I}$ scales as
\begin{equation}
\label{Eq:scaleipr}
{\mathcal I} =L^{-s/\nu}f(\varepsilon L^{1/\nu}),
\end{equation}
where $f(.)$ is the scaling function of the static $\mathcal{I}$.
In Fig.~\ref{staticIPR}, the curves of $\mathcal{I}$ as a function of $\varepsilon$ for different $L$, as well as the rescaled curves according to Eq.~\eqref{Eq:scaleipr}, are plotted.
It can be observed that all curves collapse onto a single one after rescaling, confirming the validity of Eq.~\eqref{Eq:scaleipr}.

\begin{figure}[htbp]
\centering
  \includegraphics[width=2.5in,clip]{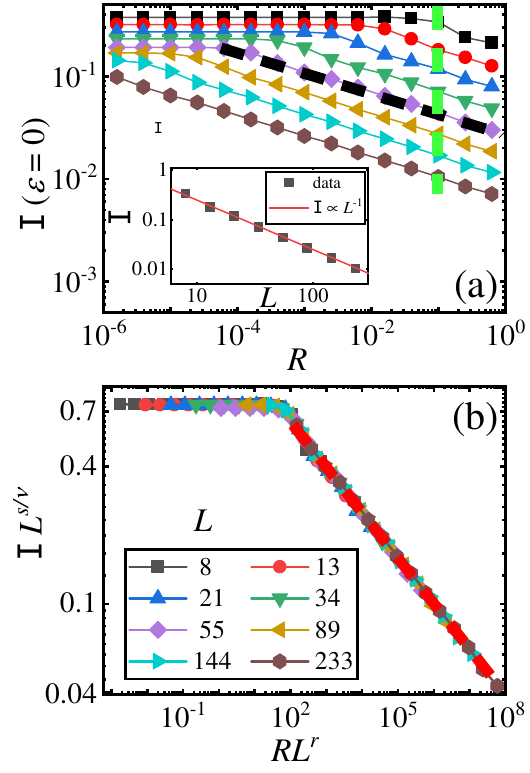}
  \vskip-3mm
  \caption{Driven dynamics of ${\mathcal I}$ with initial extended states.
 ${\mathcal I}$ at $\varepsilon=0$ versus $R$ for different $R$ (a) before and (b) after rescaling.
  For large $R$ with $L=55$ (the black dash line), power law fitting shows ${\mathcal I}\propto R^{-0.1979}$.
  Insert in (a) shows ${\mathcal I}\propto L^{-1}$ for $R=0.25$.
  In (b), the red dashed line has a slope of $-0.1982$, which is close to $(s - \nu)/(r\nu)$.
  The results are averaged for $10$ samples of $\phi$, and the log-log coordinate is used.}
  \label{IPRC}
\end{figure}

We now discuss the behavior of $\mathcal{I}$ when the system is driven to the phase transition point (i.e., $\varepsilon=0$) at different rates $R$.
In Fig.~\ref{IPRC} (a), we present ${\mathcal I}$ at $\varepsilon=0$ as a function of $R$ for different $L$.
It is shown that, for large $R$ and fixed $L$, the ${\mathcal I}$ scales as $R^{-0.1979}$ (dash black line as an example), an exponent close to $(s-\nu)/r\nu$.
For fixed large $R$, the ${\mathcal I}$ scales as $L^{-1}$ (dash green line as an example and inset).
Hence, in the large-$R$ regime we obtain
\begin{eqnarray}
 {\mathcal I} \propto L^{-1}R^{\frac{s-\nu}{r\nu}}.
\end{eqnarray}
In contrast, for small $R$, the evolution of the system will approach the equilibrium case, hence, the ${\mathcal I}$ at $\varepsilon=0$ tends to the usual finite-size scaling form of ${\mathcal I} \propto L^{-s/\nu}$.

Based on the framework of FTS~\cite{Gong2010}, the scaling form of ${\mathcal I}$ should satisfy the following relation to reconcile these scaling behaviors
\begin{eqnarray}
\label{Eq:ScalingIPR}
\mathcal{I} =\left\{ \begin{array}{l}
   L^{-\frac{s}{\nu}}f_1(RL^r,\varepsilon R^{-\frac{1}{r\nu}}), \quad \text{for small } R,  \\
  L^{-1}R^{\frac{s-\nu}{r\nu}}f_2(RL^r,\varepsilon R^{-\frac{1}{r\nu}}), \quad \text{for large } R,  \\
\end{array} \right.
\end{eqnarray}
in which $f_i(.)$ is a scaling function of driven dynamics of ${\mathcal I}$, and both $f_1$ and $f_2$
at $\varepsilon=0$ tend to become constants.
Additionally, in Eq.~\eqref{Eq:ScalingIPR}, the criteria for distinguishing between large $R$ and small $R$ vary for different $L$.
As shown in Fig.~\ref{IPRC}(a), the region where the $I$ versus $R$ curve appears as a flat straight line can be considered as small $R$, while the region where the $I$ versus $R$ curve appears as a sloping straight line can be regarded as large $R$.

Following Eq.~\eqref{Eq:ScalingIPR}, we rescale ${\mathcal I}$ and $R$ as ${\mathcal I}L^{s/\nu}$ and $RL^r$, with the results shown in Fig.~\ref{IPRC}(b).
We can see that for small $R$, the curves of ${\mathcal I}L^{s/\nu}$ versus $RL^{-r}$ for different $L$ collapse into a single flat straight line.
For large $R$, ${\mathcal I}L^{s/\nu}$ is proportional to $(RL^{-r})^{(s-\nu)/(r\nu)}$.
As a result, the rescaled curves collapse into a straight line with slope of $(s-\nu)/(r\nu)$.
These results validates the scaling relations given in Eq.~\eqref{Eq:ScalingIPR} at $\varepsilon=0$.
\begin{figure}[htbp]
\centering
  \includegraphics[width=2.3in,clip]{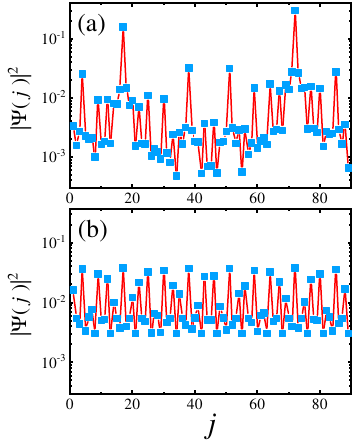}
  \vskip-3mm
  \caption{The snapshots of states during a slow ramp taken at $\varepsilon=0$.
  $R=1.6\times 10^{-6}$ in (a), $2.5\times 10^{-1}$ in (b).
  The system size is chosen as $L = 89$, with $\phi = 0.01$. The $y$-axis in the figure is plotted on a logarithmic scale.}
  \label{absPsi}
\end{figure}

Now let us discuss the form of Eq.~\eqref{Eq:ScalingIPR} at $\varepsilon=0$ from the perspective of the spatial distribution of the wave function.
As shown in Fig.~\ref{absPsi}, we present the spatial distribution of the squared modulus of the wave function $|\Psi(j)|^2$ when driven to $\varepsilon=0$ for small and large values of $R$.
It can be observed that at small $R$, the wave function displays typical features of a critical state, i.e., a self-similar structure, as shown in Fig.~\ref{absPsi} (a).
As $R$ increases, the distribution of the wave function exhibits typical characteristics of an extended state, i.e., the probability on each lattice site is relatively uniform, as shown in Fig.~\ref{absPsi} (b).
Therefore, at small $R$, ${\mathcal I}$ scaling to recover the finite-size scaling form, similar to Eq.~\eqref{Eq:scaleipr}.
At large $R$, the ${\mathcal I}$ behaves similarly to that of an extended state, exhibiting a dependence of ${\mathcal I}\propto L^{-1}$.
The remaining dimension of ${\mathcal I}$ is then borne by $R$, manifesting as a scaling relation of the form $R^{(s-\nu)/r\nu}$.

\begin{figure}[thbp]
\centering
  \includegraphics[width=3.5in,clip]{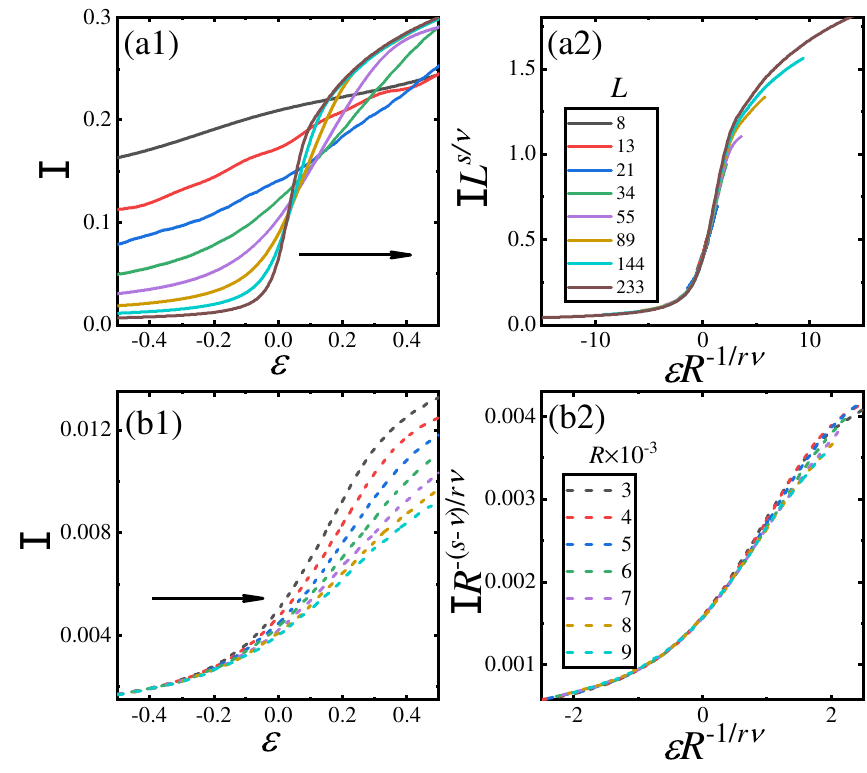}
  \vskip-3mm
  \caption{Curves of ${\mathcal I}$ versus $\varepsilon$ for fixed $RL^r=950.58$ with different $L$ (a1) before and (a2) after rescaling according to Eq.~\eqref{Eq:ScalingIPR}.
  Curves of ${\mathcal I}$ versus $\varepsilon$ for fixed $L=987$ (b1) before and (b2) after rescaling according to Eq.~\eqref{Eq:ScalingIPR3}.
  The results are averaged for $10$ samples of $\phi$.
  The arrows in (a1) and (b1) denote the quench direction.}
  \label{FixLRIPR}
\end{figure}

To further examine the applicability of FTS in the driven process, we fixed $R L^{r}$ to an arbitrary value and calculated the $\varepsilon$ dependence of the ${\mathcal I}$ for different $L$, as shown in Fig.~\ref{FixLRIPR}(a1).
Following Eq.~\eqref{Eq:ScalingIPR}, the rescaled curves of the ${\mathcal I}$ versus $\varepsilon$ collapse onto each other, as demonstrated in Fig.~\ref{FixLRIPR}(a2).
This data collapse further validates the correctness of Eq.~\eqref{Eq:ScalingIPR}.

For sufficiently large $L$, Eq.~\eqref{Eq:ScalingIPR} can be simplified to the following form
\begin{equation*}
\label{Eq:ScalingIPR2}
{{\mathcal I}}=L^{-1}R^{\frac{s-\nu}{r\nu}}f_3(\varepsilon R^{-\frac{1}{r\nu}}).
\end{equation*}
And, for fixed $L$, it becomes
\begin{equation}
\label{Eq:ScalingIPR3}
{{\mathcal I}}=R^{\frac{s-\nu}{r\nu}}f_4(\varepsilon R^{-\frac{1}{r\nu}}).
\end{equation}
As shown in Fig.~\ref{FixLRIPR}(b1) and (b2), where we fixed $L=987$ as an example, Eq.~\eqref{Eq:ScalingIPR3} was numerically verified.
The rescaled curves for different $R$ values collapse onto a universal curve when scaled according to Eq.~\eqref{Eq:ScalingIPR3}, thereby validating Eq.~\eqref{Eq:ScalingIPR3}.

When the system is driven starting from a localized state, for large $R$ the wave functions at the phase transition point still exhibit characteristics of a localized state~\cite{Sinha2019}.
Since for a localized state $\mathcal{I} \propto L^0$, the scaling equation of $\mathcal{I}$ in this case differs from that when the initial state is extended.
For sufficiently large $L$, the driven dynamics of $\mathcal{I}$ with a initial localized state satisfies,
\begin{equation}
\label{Eq:ScalingIPRlocal}
{{\mathcal I}}=R^{\frac{s}{r\nu}}f_5(\varepsilon R^{-\frac{1}{r\nu}}).
\end{equation}
Eq.~\eqref{Eq:ScalingIPRlocal} has been confirmed in multiple studies~\cite{Zhai2022,Sinha2019}.

\subsection{Driven dynamics of $\mathcal{D}$}

The dynamic deviation from the instantaneous ground state energy $\mathcal D$ is also an important quantity in characterizing the dynamical behavior of localization phase transitions~\cite{Zhai2022}.
Its definition is given by
\begin{equation}
\label{DEG}
\mathcal{D}(t) = \Psi^*(t)  H(t) \Psi(t) - E_g(t),
\end{equation}
where $H(t)$ is the instantaneous Hamiltonian in the driving process, and $E_g(t)$ is the corresponding ground state energy. $\Psi(t)=e^{iH(t)}\Psi(0)$, and $\Psi(0)$ is the initial state.
As can be seen from Eq.~\eqref{DEG}, $\mathcal{D}$ always vanishes at $t=0$, but it shares the same dimension as $\Delta E$.

\begin{figure}[htbp]
\centering
  \includegraphics[width=2.5in,clip]{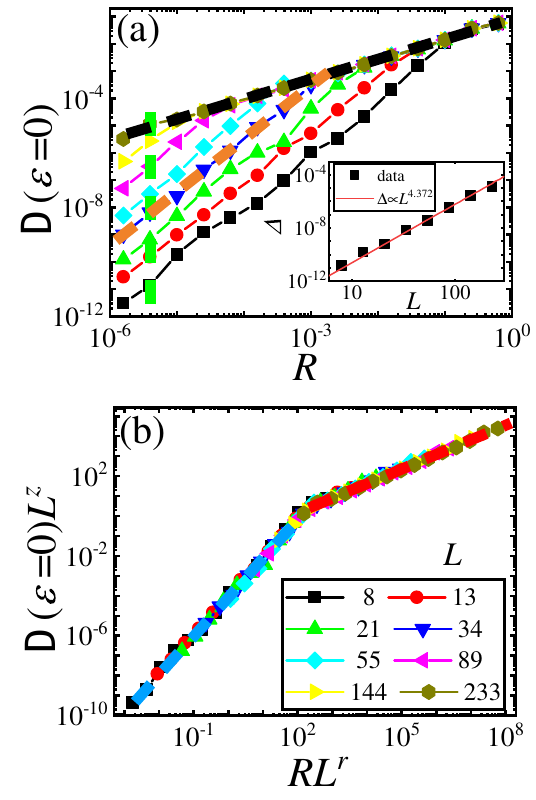}
  \vskip-3mm
  \caption{Driven dynamics of $\mathcal{D}$ with initial extended states.
 ${\mathcal D}$ at $\varepsilon=0$ versus $R$ for different $R$ (a) before and (b) after rescaling.
  In (a), for small $R$ with $L=34$, power law fitting shows ${\mathcal D}\propto R^{2}$ (the orange dash line as an example).
  Insert shows ${\mathcal D}\propto L^{4.372}$ for $R=3.98\times 10^{-6}$ (the green dash line).
  For large $R$, power law fitting shows ${\mathcal D}\propto R^{0.7033}$ (the black dash line).
  In (b), the slops of the blue and red dash lines are $1.998$ and $0.7045$.
  The results are averaged for $10$ samples of $\phi$, and the log-log coordinate is used.}
  \label{EGC}
\end{figure}
As shown in Fig.~\ref{EGC}(a), we present the $\mathcal{D}$ at $\varepsilon=0$ versus $R$ curves for different $L$.
We find that for small $R$, the dependence of $\mathcal{D}$ on $R$ for a fixed $L$ follows $\mathcal{D}\propto R^2$ (the orange dashed line as an example); while for a fixed $R$, the dependence of $\mathcal{D}$ on $L$ exhibits $\mathcal{D}\propto L^{4.37}$ (see the inset of Fig.~\ref{EGC}(a)), where the exponent is close to $2r-z$.
Consequently, in the small-$R$ regime, we can write
\begin{equation}
\label{DEGscale}
\mathcal{D} \propto R^2L^{2r-z}.
\end{equation}
It is important to emphasize that this scaling behavior, where $\mathcal{D}$ is proportional to $R^2$, is reported here for the first time.
For large $R$, we observe that $\mathcal{D}$ scales as $R^{0.7033}$ and is independent of $L$, with this exponent closely matching the value of $z/r$, as indicated by the black dashed line in Fig.~\ref{EGC}(a).

To construct a unified scaling equation that simultaneously satisfies the aforementioned scaling relations while incorporating both finite-size effects and off-critical-point corrections, we propose the following scaling ansatz
\begin{eqnarray}
\label{Eq:ScalingDEG}
\mathcal{D} =\left\{ \begin{array}{l}
  {{R}^{2}}{{L}^{2r-z}}g_1(R{{L}^{r}},\varepsilon R^{-\frac{1}{r\nu}}), \quad \text{for small } R,  \\
  {{R}^{\frac {z}{r}}}g_2(R{{L}^{r}},\varepsilon R^{-\frac{1}{r\nu}}) , \quad \text{for large } R,  \\
\end{array} \right.
\end{eqnarray}
where $g_i(.)$ is the scaling function for the driven dynamics of $\mathcal{D}$.
At $\varepsilon=0$, $g_1$ and $g_2$ tend to become constants.

\begin{figure}[htbp]
\centering
  \includegraphics[width=3.5in,clip]{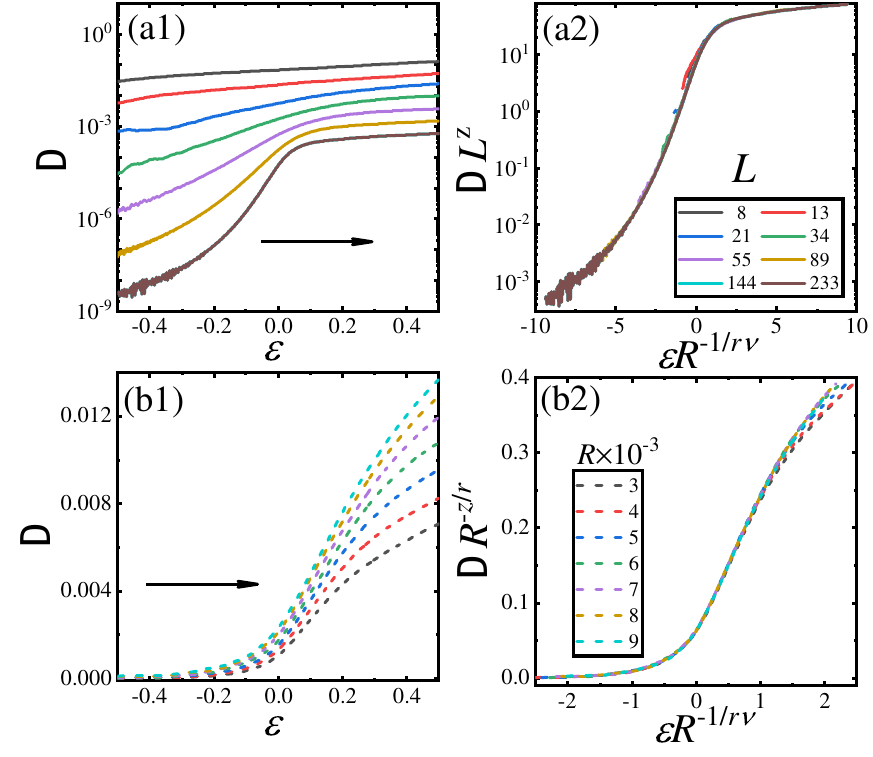}
  \vskip-3mm
  \caption{Curves of $\mathcal{D}$ versus $\varepsilon$ for fixed $RL^r=950.58$ with different $L$ (a1) before and (a2) after rescaling according to Eq.~\eqref{Eq:ScalingIPR}.
  Curves of $\Delta$ versus $\varepsilon$ for fixed $L=987$ (b1) before and (b2) after rescaling according to Eq.~\eqref{Eq:ScalingIPR3}.
  The results are averaged for $10$ samples of $\phi$.
  The arrows in (a1) and (b1) denote the quench direction.}
  \label{FixLREG}
\end{figure}

To verify this scaling equation, we rescale $\mathcal{D}$ at $\varepsilon=0$ and $R$ as $\mathcal{D}L^z$ and $RL^r$.
According to Eq.~\eqref{Eq:ScalingDEG}, $\mathcal{D}L^z$ scales as $(RL^r)^2$ for small $R$, while for large $R$, $\mathcal{D}L^z\propto(RL^r)^{z/r}$.
As shown in Fig.~\ref{EGC}(b), the rescaled curves collapse into a straight line with a slope close to 2 in the small-$R$ region, while in the large-$R$ region, they collapse into another straight line with a slope close to 0.7033.
This data collapse further confirms the validity of Eq.~\eqref{Eq:ScalingDEG} at $\varepsilon=0$.

We now discuss the origin of this scaling form for $\mathcal{D}$ at $\varepsilon=0$.
In the small-$R$ region, although the gap of extended states becomes vanishingly small in finite systems, the initial state can still be considered gapped.
Consequently, the gap dominates the driving dynamics.
For fixed $L$, $\mathcal{D}$ can be expanded as $\mathcal{D} = \sum_{n} A_n R^n$.
Since $\mathcal{D}$ must remain positive regardless of whether $R$ is positive or negative, the expansion should contain only even powers of $R$.
Keeping the lowest order term, we obtain $\mathcal{D} \propto R^2$.
Concurrently, the remaining scaling dimension must be borne by $L$, which gives rise to $\mathcal{D} \propto L^{2r - z}$.
In the large-$R$ regime, as $\mathcal{D}$ shares the same dimension as the $\Delta E$ and the driven dynamics at large $R$ can be described by conventional finite-size effects, we obtain $\mathcal{D} \propto R^{z/r}$.

To further validate the correctness of Eq.~\eqref{Eq:ScalingDEG} during the driving dynamics, we fix $RL^{r}$ to an arbitrary constant value.
As shown in Fig.~\ref{FixLREG}(a1), we present the $\mathcal{D}$ versus $\varepsilon$ curves for different $L$ when $RL^{r}=950.58$.
After rescaling, all curves collapse onto a single universal curve as shown in Fig.~\ref{FixLREG}(a2), confirming the validity of Eq.~\eqref{Eq:ScalingDEG}.

For sufficiently large $L$, Eq.~\eqref{Eq:ScalingDEG} can be simplified to
\begin{equation}
\label{Eq:ScalingDEG3}
\mathcal{D} = R^{\frac{z}{r}} g_3(\varepsilon R^{-\frac{1}{r\nu}}).
\end{equation}
In Fig.~\ref{FixLREG}(b1), we show the $\mathcal{D}$ versus $\varepsilon$ curves for different $R$ values at $L=987$.
The rescaled curves exhibit perfect collapse as shown in Fig.~\ref{FixLREG}(b2), thereby verifying Eq.~\eqref{Eq:ScalingDEG3}.
Note that Eq.~\eqref{Eq:ScalingDEG3} is identical in form to the scaling equation for the initial localized state~\cite{Sinha2019,Zhai2022}, since $\mathcal{D}$ is always zero at $t=0$.

\section{\label{sum}conclusion}
In summary, we have systematically studied the driven dynamics of the localization phase transition in the AA model starting from a gapless extended state.
Through extensive numerical simulations and scaling analysis, we have demonstrated that the driven dynamics in this regime is well described by the FTS framework.
We have established that the driven dynamics with initial extended states satisfies the FTS applicability criterion ($z' < r$), enabling a complete scaling description of the full non-equilibrium evolution.
The scaling functions for both the ${\mathcal I}$ and ${\mathcal D}$ have been derived and numerically validated.
We observe that the scaling functions of these order parameters differ significantly between the small-$R$ and large-$R$ regimes.
In particular, we report for the first time that ${\mathcal D}$ exhibits an $R^2$ dependence in the small-$R$ region.
Furthermore, these scaling relations exhibit fundamentally different forms compared to those derived from localized initial states, highlighting the crucial role of initial state behavior in the ensuing driven dynamics.
The robustness of these scaling functions across various system sizes and driving rates underscores the universal nature of the non-equilibrium dynamics originating from gapless extended states.

Our work not only extends the understanding of driven dynamics in localization phase transitions beyond the conventional initial localized state, but also highlights the significant influence of initial state topology on the resulting non-equilibrium behavior. Given experimental advances in realizing the AA model with ultracold atoms and optical lattice~\cite{Roati2008,Billy2008}, the proposed scaling relations provide directly testable predictions for future experimental investigations of non-equilibrium dynamics in quasiperiodic systems.
\section*{Acknowledgments}
This work is supported by the National Natural Science Foundation of China (Grant Nos. 12274184 and 12404105), the Qing Lan Project, and the Natural Science Foundation of the Jiangsu Higher Education Institutions of China (Grant No. 24KJB140008).


\begin{thebibliography}{99}
\bibitem{Anderson1958}P. W. Anderson, Absence of diffusion in certain random lattices, Phys. Rev. {\bf109}, 1492 (1958).
\bibitem{Abrahams1979}E. Abrahams, P.W. Anderson, D.C. Licciardello, and T.V. Ramakrishnan, Scaling Theory of Localization: Absence of Quantum Diffusion in Two Dimensions, Phys. Rev. Lett. {\bf 42}, 673-676 (1979).
\bibitem{Lee1985}P. A. Lee and T. V. Ramakrishnan, Disordered electronic systems, Rev. Mod. Phys. {\bf57}, 287 (1985).
\bibitem{Kramer1993}B. Kramer, A. MacKinnon, Localization: theory and experiment, Rep. Prog. Phys. {\bf56}, 1469 (1993).
\bibitem{Schwartz2007}T. Schwartz, G. Bartal, S. Fishman, and M. Segev, Transport and Anderson localization in disordered two-dimensional photonic lattices, Nature {\bf446}, 7131 (2007).
\bibitem{Evers2008}F. Evers, A.D. Mirlin, Anderson transitions, Rev. Mod. Phys., {\bf80}, 1355-1417 (2008).
\bibitem{Segev2013}M. Segev, Y. Silberberg, and D. N. Christodoulides, Anderson localization of light, Nature Photonics {\bf7}, 197 (2013).
\bibitem{Wiersma2013}D. S. Wiersma, Disordered photonics, Nature Photonics {\bf7}, 188 (2013).
\bibitem{Alexey2023}A. Yamilov, S. E. Skipetrov, T. W. Hughes, M. Minkov, Z. Yu, and H. Cao, Anderson localization of electromagnetic waves in three dimensions, Nat. Phys. {\bf19}, 1308 (2023).


\bibitem{Aubry1980}S. Aubry and G. Andr\'{e}, Analyticity breaking and anderson localization in incommensurate lattices, Ann. Israel Phys. Soc. {\bf3}, 133 (1980).
\bibitem{Biddle070601}J. Biddle and S. Das Sarma, Predicted mobility edges in one-dimensional incommensurate optical lattices: An exactly solvable model of anderson localization, Phys. Rev. Lett. {\bf104}, 070601 (2010).
\bibitem{HepengYao2019}H. Yao, A. Khoudli, L. Bresque, and L. Sanchez-Palencia, Critical behavior and fractality in shallow one-dimensional quasiperiodic potentials, Phys. Rev. Lett. {\bf123}, 070405 (2019).
\bibitem{Schirmann2024}J. Schirmann, S. Franca, F. Flicker, and A. G. Grushin, Physical Properties of an Aperiodic Monotile with Graphene-like Features, Chirality, and Zero Modes, Phys. Rev. Lett. {\bf132}, 086402 (2024).
\bibitem{Ribeiro2024}M. Gon\c{c}alves, B. Amorim, E. V. Castro, and P. Ribeiro, Critical Phase Dualities in 1D Exactly Solvable Quasiperiodic Models, Phys. Rev. Lett. {\bf131}, 186303 (2024).
\bibitem{Jan2023}J. \v{S}untajs, T. c. v. Prosen, and L. Vidmar, Localization challenges quantum chaos in the finite two-dimensional Anderson model, Phys. Rev. B {\bf107}, 064205 (2023).
\bibitem{Agrawal2020}U. Agrawal, S. Gopalakrishnan, and R. Vasseur, Universality and quantum criticality in quasiperiodic spin chains, Nat. Commun. {\bf11}, 2225 (2020).
\bibitem{Agrawalprl2020}U. Agrawal, S. Gopalakrishnan, and R. Vasseur, Quantum Criticality in the 2D Quasiperiodic Potts Model, Phys. Rev. Lett. {\bf125}, 265702 (2020).
\bibitem{Goblot2020}V. Goblot, A. \v{S}trkalj, N. Pernet, J. L. Lado, C. Dorow, A. Lema\^{\i}tre, L. L. Gratiet, A. Harouri, I. Sagnes, S. Ravets, A. Amo, J. Bloch, and O. Zilberberg, Emergence of criticality through a cascade of delocalization transitions in quasiperiodic chains, Nat. Phys. {\bf16}, 832 (2020).
\bibitem{Roy2022}S. Roy, S. Chattopadhyay, T. Mishra, and S. Basu, Critical analysis of the reentrant localization transition in a one-dimensional dimerized quasiperiodic lattice, Phys. Rev. B {\bf105}, 214203 (2022).
\bibitem{Agrawal2022}U. Agrawal, R. Vasseur, and S. Gopalakrishnan, Quasiperiodic many-body localization transition in dimension $d\textgreater1$, Phys. Rev. B {\bf106}, 094206 (2022).

\bibitem{Zhou2023}X.-C. Zhou, Y.-J. Wang, T.-F. J. Poon, Q. Zhou, and X.-J. Liu, Exact New Mobility Edges between Critical and Localized States, Phys. Rev. Lett. {\bf131}, 176401 (2023).
\bibitem{Smith176601}A. Smith, J. Knolle, R. Moessner, D. L. Kovrizhin, Absence of ergodicity without quenched disorder: From quantum disentangled liquids to many-body localization, Phys. Rev. Lett. {\bf119}, 176601 (2017).

\bibitem{Biddle2011}J. Biddle, D. J. Priour, B. Wang, and S. Das Sarma, Localization in one-dimensional lattices with non-nearest-neighbor hopping: Generalized Anderson and Aubry-Andr\'{e} models, Phys. Rev. B {\bf83}, 075105 (2011).
\bibitem{Kraus2012}Y. E. Kraus, Y. Lahini, Z. Ringel, M. Verbin, and O. Zilberberg, Topological States and Adiabatic Pumping in Quasicrystals, Phys. Rev. Lett. {\bf109}, 106402 (2012).
\bibitem{Xuan2022}X. Bu, L.-J. Zhai, and S. Yin, Quantum criticality in the disordered Aubry-Andr\'{e} model, Phys. Rev. B {\bf106}, 214208 (2022).
\bibitem{Ting2022}T. Lv, Y.-B. Liu, T.-C. Yi, L. Li, M. Liu, and W.-L. You, Exploring unconventional quantum criticality in the $p$-wave-paired Aubry-Andr\'{e}-Harper model, Phys. Rev. B {\bf106}, 144205 (2022).
\bibitem{Sahoo2025}A. Sahoo, A. Saha, and D. Rakshit, Stark localization near Aubry-Andr\'{e} criticality, Phys. Rev. B {\bf 111}, 024205 (2025).
\bibitem{Guo2024}C. Y. Guo, Multiple intermediate phases in the interpolating Aubry-Andr\'{e}-Fibonacci model, Phys. Rev. B {\bf109}, 174203 (2024).
\bibitem{Yi2025}T.-C. Yi, Y.-Y. Fang, W. Chen, W.-L. You, and Y. Zhang, Unveiling quantum criticality of disordered Aubry-Andr\'e-Harper models via typical fidelity susceptibility, Phys. Rev. A {\bf112}, 023308 (2025).


\bibitem{Saha2016}S. Saha, S. K. Maiti, and S. Karmakar, Multiple mobility edges in a 1D Aubry chain with Hubbard interaction in presence of
electric field: Controlled electron transport, Physica E {\bf83}, 358 (2016).
\bibitem{Ganeshan2015}S. Ganeshan, J. H. Pixley, and S. Das Sarma, Nearest neighbor tight binding models with an exact mobility edge in one dimension, Phys. Rev. Lett. {\bf114}, 146601 (2015).
\bibitem{Wangprl2020}Y. Wang, X. Xia, L. Zhang, H. Yao, S. Chen, J. You, Q. Zhou, and X.-J. Liu, One-dimensional quasiperiodic mosaic lattice with exact mobility edges, Phys. Rev. Lett. {\bf125}, 196604 (2020).
\bibitem{Hu2025}H.-T. Hu, Y. Chen, X. Lin, A.-M. Guo, Z. Lin, and M. Gong, Exact mobility edges in quasiperiodic network models with slowly varying potentials, Phys. Rev. B {\bf112}, 054201 (2025).
\bibitem{Liu2022}T. Liu, X. Xia, S. Longhi, and L. Sanchez-Palencia, Anomalous mobility edges in one-dimensional quasiperiodic models, SciPost Phys. {\bf12}, 027 (2022).
\bibitem{Gao2024}J. Gao, Ivan M. Khaymovich, X.-W. Wang, Z.-S. Xu, A. Iovan, G. Krishna, J. Jieensi, A. Cataldo, A. V. Balatsky, V. Zwiller, and A. W. Elshaari, Probing multi-mobility edges in quasiperiodic mosaic lattices, Sci. Bull. {\bf09}, 030 (2024).
\bibitem{Banerjee2025}S. Banerjee, S. R. Padhi, T. Mishra, Emergence of distinct exact mobility edges in a quasiperiodic chain, arXiv:2503.19834 (2025).

\bibitem{Iyer2013}S. Iyer, V. Oganesyan, G. Refael, and D. A. Huse, Many-body localization in a quasiperiodic system, Phys. Rev. B {\bf87}, 134202 (2013).
\bibitem{Schreiber2015}M. Schreiber, S. S. Hodgman, P. Bordia, H. P. L\"{u}schen, M. H. Fischer, R. Vosk, E. Altman, U. Schneider, and I. Bloch, Observation of many-body localization of interacting fermions in a quasirandom optical lattice, Science {\bf349}, 842 (2015).
\bibitem{Mastropietro2015}V. Mastropietro, Localization of Interacting Fermions in the Aubry-Andr\'e Model, Phys. Rev. Lett. {\bf115}, 180401 (2015).
\bibitem{Khemani2017}V. Khemani, D. N. Sheng, and D. A. Huse, Two universality classes for the many-body localization transition, Phys. Rev. Lett. {\bf119}, 075702 (2017).
\bibitem{Zhang2018}S.-X. Zhang and H. Yao, Universal Properties of Many-Body Localization Transitions in Quasiperiodic Systems, Phys. Rev. Lett. {\bf121}, 206601 (2018).
\bibitem{Rispoli2019}M. Rispoli, A. Lukin, R. Schittko, S. Kim, M. E. Tai, J. L\'{e}onard, and M. Greiner, Quantum critical behaviour at the many-body localization transition, Nature {\bf573}, 385 (2019).
\bibitem{Weiner2019}F. Weiner, F. Evers, and S. Bera, Slow dynamics and strong finite-size effects in many-body localization with random and quasiperiodic potentials, Phys. Rev. B {\bf100}, 104204 (2019).
\bibitem{Xu2019}S. Xu, X. Li, Y.-T. Hsu, B. Swingle, and S. Das Sarma, Buttery effect in interacting Aubry-Andr\'{e} model: Thermalization, slow scrambling, and many-body localization, Phys. Rev. Research {\bf1}, 032039 (2019).
\bibitem{zhai2020}L.-J. Zhai, S. Yin, and G.-Y. Huang, Many-body localization in a non-hermitian quasiperiodic system, Phys. Rev. B {\bf102}, 064206 (2020).
\bibitem{Cookmeyer2020}T. Cookmeyer, J. Motruk, and J. E. Moore, Critical properties of the ground-state localization-delocalization transition in the many-particle Aubry-Andr\'e model, Phys. Rev. B {\bf101}, 174203 (2020).
\bibitem{Strkalj2021}A. \v{S}trkalj, E. V. H. Doggen, I. V. Gornyi, and O. Zilberberg, Many-body localization in the interpolating Aubry-Andr\'{e}-Fibonacci model, Phys. Rev. Research {\bf3}, 033257 (2021).
\bibitem{Tang2021}L.-Z. Tang, G.-Q. Zhang, L.-F. Zhang, D.-W. Zhang, Localization and topological transitions in non-Hermitian quasiperiodic lattices, Phys. Rev. A, {\bf103}, 033325 (2021).

\bibitem{Strkalj2022}A. \v{S}trkalj, E.V.H. Doggen, C. Castelnovo, Coexistence of localization and transport in many-body two-dimensional Aubry-Andr\'e models, Phys. Rev. B, {\bf106}, 184209 (2022).
\bibitem{Prasad2024}Y. Prasad and A. Garg, Single-particle excitations across the localization and many-body localization transition in quasiperiodic systems, Phys. Rev. B {\bf109}, 094204 (2024).
\bibitem{Huang2024}K. Huang, D. Vu, S. Das Sarma, X. Li, Interaction-enhanced many-body localization in a generalized Aubry-Andr\'e model, Phys. Rev. Research, {\bf6}, L022054 (2024).
\bibitem{Chakrabarty2025}A. Chakrabarty, S. Banerjee, T. Mishra, and S. Datta, Emergence of non-trivial phases in interacting non-Hermitian quasiperiodic chains with power-law hopping, arXiv:2508.14724 (2025).

\bibitem{Zeng2017}Q.-B. Zeng, S. Chen, R. L\"{u}, Anderson localization in the non-Hermitian Aubry-Andr\'e-Harper model with physical gain and loss, Phys. Rev. A, {\bf95}, 062118 (2017).
\bibitem{Jiang2019}H. Jiang, L.-J. Lang, C. Yang, S.-L. Zhu, and S. Chen, Interplay of non-hermitian skin effects and Anderson localization in nonreciprocal quasiperiodic lattices, Phys. Rev. B {\bf100}, 054301 (2019).
\bibitem{Longhi20192}S. Longhi, Metal-insulator phase transition in a non-Hermitian Aubry-Andr\'{e}-Harper model, Phys. Rev. B {\bf100}, 125157 (2019).
\bibitem{Longhi224206}S. Longhi, Non-Hermitian maryland model, Phys. Rev. B {\bf103}, 224206 (2021).
\bibitem{Liu2021}Y. Liu, Y. Wang, X.-J. Liu, Q. Zhou, and S. Chen, Exact mobility edges, PT-symmetry breaking, and skin effect in one-dimensional non-hermitian quasicrystals, Phys. Rev. B {\bf103}, 014203 (2021).
\bibitem{Guo2021}C.-X. Guo, C.-H. Liu, X.-M. Zhao, Y. Liu, and S. Chen, Exact solution of non-Hermitian systems with generalized boundary conditions: Size-dependent boundary effect and fragility of the skin effect, Phys. Rev. Lett. {\bf127}, 116801 (2021).
\bibitem{zhai2021}L.-J. Zhai, G.-Y. Huang, and S. Yin, Cascade of the delocalization transition in a non-Hermitian interpolating Aubry-Andr\'{e}-Fibonacci chain, Phys. Rev. B {\bf104}, 014202 (2021).
\bibitem{xueprl2022}Q. Lin, T.-Y Li, L. Xiao, K. -K Wang, W. Yi, P. Xue, Topological Phase Transitions and Mobility Edges in Non-Hermitian Quasicrystals, Phys. Rev. Lett. {\bf129}, 113601 (2022).
\bibitem{Acharya2024}A. P. Acharya and S. Datta, Localization transitions in a non-Hermitian quasiperiodic lattice, Phys. Rev. B {\bf109}, 024203 (2024).
\bibitem{Guocx2024}C.-X. Guo, L. H. Su, Y. L. Wang, L. Li, J. Z. Wang, X. H Ruan, Y. J. Du, D. N. Zheng, S. Chen, and H. P Hu, Scale-tailored localization and its observation in non-Hermitian electrical circuits, Nat. Commun. {\bf15}, 9120 (2024).
\bibitem{Zeng2024}C.-C. Zeng, Z. Cai, G.-H. Wang, G. Sun, Fidelity and criticality in the nonreciprocal Aubry-Andr\'{e}-Harper model, Europhys. Lett., {\bf149}, 38001 (2025).
\bibitem{Sun2024}Y.-M. Sun, X.-Y. Wang and L.-J. Zhai, Hybrid scaling properties of the localization transition in a non-Hermitian disordered Aubry-Andr\'{e} model, Phys. Rev. B {\bf110}, 054202 (2024).
\bibitem{Chen2025}R.-J. Chen, G.-Q. Zhang, Z. Li, D.-W. Zhang, Mobility rings in a non-Hermitian non-Abelian quasiperiodic lattice, Phys. Rev. A, {\bf112}, 013320 (2025).

\bibitem{Blatt2012}R. Blatt and C. F. Roos, Quantum simulations with trapped ions, Nature Phys. {\bf8}, 277 (2012).
\bibitem{Mukherjee2017}S. Mukherjee, A. Spracklen, M. Valiente, E. Andersson, P. \"{O}hberg, N. Goldman, and R. R. Thomson, Experimental observation of anomalous topological edge modes in a slowly driven photonic lattice, Nature Commun. {\bf8}, 13918 (2017).
\bibitem{Gross2017}C. Gross and I. Bloch, Quantum simulations with ultracold atoms in optical lattices, Science {\bf357}, 995 (2017).
\bibitem{Ebadi2021}S. Ebadi, T. T. Wang, H. Levine, A. Keesling, G. Semeghini, A. Omran, D. Bluvstein, R. Samajdar, H. Pichler, W. W. Ho, S. Choi, S. Sachdev, M. Greiner, V. Vuleti\'{c}, M. D. Lukin, Quantum phases of matter on a 256-atom programmable quantum simulator, Nature {\bf595}, 227-232 (2021).
\bibitem{xueKZM2021} L. Xiao, D. -K. Qu, K. -K. Wang, H. -W Li, J.-Y Dai, B. D\'{o}ra , M. Heyl, R. Moessner, W. Yi, and P. Xue, Non-Hermitian Kibble-Zurek Mechanism with Tunable Complexity in Single-Photon Interferometry, PRX QUANTUM {\bf2}, 020313 (2021).


\bibitem{Bedow2022}J. Bedow, E. Mascot, and D. K. Morr, Emergence and manipulation of non-equilibrium Yu-Shiba-Rusinov states. Commun. Phys. {\bf5}, 281 (2022).
\bibitem{Perfetto2021}G. Perfetto, F. Carollo, M. Magoni, I. Lesanovsky, Designing nonequilibrium states of quantum matter through stochastic resetting, Phys. Rev. B {\bf104}, L180302 (2021).
\bibitem{Mankowsky2016}R. Mankowsky, M. F\"{o}rst, A. Cavalleri, Non-equilibrium control of complex solids by nonlinear phononics, Rep. Prog. Phys. {\bf79}, 064503 (2016).
\bibitem{Popkov2013}V. Popkov, R. Livi, Manipulating energy and spin currents in non-equilibrium systems of interacting qubits, New J. Phys. {\bf15}, 023030 (2013).
    \bibitem{Giudici2022}G. Giudici, M.D. Lukin, and H. Pichler, Dynamical Preparation of Quantum Spin Liquids in Rydberg Atom Arrays, Phys. Rev. Lett. {\bf129}, 090401 (2022).
\bibitem{Cheng2023}Y. Cheng, C. Li, and H. Zhai, Variational approach to quantum spin liquid in a Rydberg atom simulator, New J. Phys., {\bf25}, 033010 (2023).

\bibitem{Keesling2019}A. Keesling, A. Omran, H. Levine, H. Bernien, H. Pichler, S. Choi, R. Samajdar, S. Schwartz, P. Silvi, S. Sachdev, P. Zoller, M. Endres, M. Greiner, V. Vuleti\'{c}, and M. D. Lukin, Quantum Kibble-Zurek mechanism and critical dynamics on a programmable Rydberg simulator, Nature {\bf 568}, 207 (2019).
\bibitem{Dupont2022}M. Dupont and J. E. Moore, Quantum criticality using a superconducting quantum processor, Phys. Rev. B {\bf 106}, L041109 (2022).
\bibitem{Reichhardt2022}C.J.O. Reichhardt, A. del Campo, C. Reichhardt, Kibble-Zurek mechanism for nonequilibrium phase transitions in driven systems with quenched disorder, Commun. Phys. {\bf5}, 173 (2022).
\bibitem{King2023}A. D. King, J. Raymond, T. Lanting, R. Harris, A. Zucca, F. Altomare, A. J. Berkley, K. Boothby, S. Ejtemaee, C. Enderud, E. Hoskinson, S. Huang, E. Ladizinsky, A. J. R. MacDonald, G. Marsden, R. Molavi, T. Oh, G. Poulin-Lamarre, M. Reis, C. Rich, Y. Sato, N. Tsai, M. Volkmann, J. D. Whittaker, J. Yao, A. W. Sandvik, and M. H. Amin, Quantum critical dynamics in a 5,000-qubit programmable spin glass, Nature {\bf 617}, 61 (2023).
\bibitem{Soto2024}J. Soto Garcia and N. Chepiga, Resolving chiral transitions in one-dimensional Rydberg arrays with quantum Kibble-Zurek mechanism and finite-time scaling, Phys. Rev. B {\bf 110}, 125113 (2024).



\bibitem{Kibble1976}T. W. B. Kibble, Topology of cosmic domains and strings, J. Phys. A: Math. Gen. {\bf 9}, 1387 (1976).
\bibitem{Zurek1985}W. H. Zurek, Cosmological experiments in superfluid helium?, Nature {\bf 317}, 505 (1985).
\bibitem{Zurek2005}W. H. Zurek, U. Dorner, and P. Zoller, Dynamics of a quantum phase transition, Phys. Rev. Lett. {\bf 95}, 105701 (2005).
\bibitem{Dziarmaga2005}J. Dziarmaga, Dynamics of a quantum phase transition: exact solution of the quantum Ising model, Phys. Rev. Lett. {\bf 95}, 245701 (2005).
\bibitem{Dziarmaga2010}J. Dziarmaga, Dynamics of a quantum phase transition and relaxation to a steady state, Adv. Phys. {\bf59}, 1063 (2010).
\bibitem{Polkovnikov2011}A. Polkovnikov, K. Sengupta, A. Silva, and M. Vengalattore, Colloquium: nonequilibrium dynamics of closed interacting quantum systems, Rev. Mod. Phys. {\bf83}, 863 (2011).
\bibitem{Yin2017}S. Yin, G.-Y. Huang, C.-Y. Lo, and P. Chen, Kibble-Zurek scaling in the Yang-Lee edge singularity, Phys. Rev. Lett. {\bf118}, 065701 (2017).
\bibitem{zhai2018}L.-J. Zhai, H.-Y. Wang, and S. Yin, Hybridized Kibble-Zurek scaling in the driven critical dynamics across an overlapping critical region, Phys. Rev. B {\bf97}, 134108 (2018).
\bibitem{Shu2025}Y.-R. Shu, S.-K. Jian, A. W. Sandvik, and S. Yin, Equilibration of topological defects near the deconfined quantum multicritical point, Nat. Commun. {\bf16}, 3402 (2025).

\bibitem{Deng2008}S. Deng, G. Ortiz, and L. Viola, Dynamical non-ergodic scaling in continuous finite-order quantum phase transitions, Europhys. Lett. {\bf84}, 67008 (2008).
\bibitem{Polkovnikov2008}A. Polkovnikov and V. Gritsev, Breakdown of the adiabatic limit in low-dimensional gapless systems, Nat. Phys. {\bf4}, 477 (2008).
\bibitem{Pellegrini2008}F. Pellegrini, S. Montangero, G. E. Santoro, and R. Fazio, Adiabatic quenches through an extended quantum critical region, Phys. Rev. B {\bf77}, 140404 (2008).

\bibitem{Gong2010}S. Gong, F. Zhong, X. Huang, and S. Fan, Finite-time scaling via linear driving, New J. Phys. {\bf12}, 043036 (2010).
\bibitem{Huang2014}Y. Huang, S. Yin, B. Feng, and F. Zhong, Kibble-Zurek mechanism and finite-time scaling, Phys. Rev. B {\bf90}, 134108 (2014).
\bibitem{Feng2016}B. Feng, S. Yin, and F. Zhong, Theory of driven nonequilibrium critical phenomena, Phys. Rev. B {\bf94}, 144103 (2016).

\bibitem{Zeng2025}Z. Zeng, Y.-K. Yu, Z.-X. Li, Z.-X. Li, and S. Yin, Finite-time scaling beyond the Kibble-Zurek prerequisite in Dirac systems, Nature Communications {\bf16}, 6181 (2025).

\bibitem{Morales2014}L. Morales-Molina, E. Doerner, C. Danieli, and S. Flach, Resonant extended states in driven quasiperiodic lattices: Aubry-Andre localization by design, Phys. Rev. A {\bf 90}, 043630 (2014).
\bibitem{Bairey2017}E. Bairey, G. Refael, and N. H. Lindner, Driving induced many-body localization, Phys. Rev. B {\bf 96}, 020201 (2017).
\bibitem{Zhou2021}L. Zhou, Floquet engineering of topological localization transitions and mobility edges in one-dimensional non-Hermitian quasicrystals, Phys. Rev. Research {\bf 3}, 033184 (2021).
\bibitem{Tiwari2024}V. Tiwari, D. S. Bhakuni, and A. Sharma, Dynamical localization and slow dynamics in quasiperiodically driven quantum systems, Phys. Rev. B {\bf109}, L161104 (2024).
\bibitem{Yang2017}C. Yang, Y. C. Wang, P. Wang, X. L. Gao, and S. Chen, Dynamical signature of localization-delocalization transition in a one-dimensional incommensurate lattice, Phys. Rev. B {\bf95}, 184201 (2017).
\bibitem{Xu2020}Z. Xu, H. Huangfu, Y. Zhang, S. Chen, Dynamical observation of mobility edges in one-dimensional incommensurate optical lattices, New J. Phys. {\bf22}, 013036 (2020).
\bibitem{Xuzh2021}Z. H. Xu and S. Chen, Dynamical evolution in a one-dimensional incommensurate lattice with PT symmetry, Phys. Rev. A {\bf103}, 043325 (2021).
\bibitem{Cheng2399}J.-Q. Cheng, S. Yin, and D.-X. Yao, Dynamical localization transition in the non-Hermitian lattice gauge theory, Commun. Phys. {\bf7}, 58 (2024).

\bibitem{Sinha2019}A. Sinha, M. M. Rams, and J. Dziarmaga, Kibble-zurek mechanism with a single particle: Dynamics of the localization-delocalization transition in the Aubry-Andr\'{e} model, Phys. Rev. B {\bf99}, 094203 (2019).
\bibitem{Zhai2022}L.-J. Zhai, G.-Y. Huang, and S. Yin, Nonequilibrium dynamics of the localization-delocalization transition in the non-Hermitian Aubry-Andr\'{e} model, Phys. Rev. B {\bf106}, 014204 (2022).
\bibitem{Zhaifphy}L.-J. Zhai, L.-L. Hou, Q. Gao, and H.-Y. Wang, Kibble-Zurek scaling of the dynamical localization-skin effect phase transition in a non-Hermitian quasi-periodic system under the open boundary condition, Front. Phys. {\bf10}, 1098551 (2022).
\bibitem{Xuan2023}X. Bu, L.-J. Zhai, and S. Yin, Kibble-Zurek scaling in one-dimensional localization transitions, Phys. Rev. A {\bf108}, 023312 (2023).
\bibitem{Liang2024}E.-W. Liang, L.-Z. Tang, and D.-W. Zhang, Quantum criticality and Kibble-Zurek scaling in the Aubry-Andr\'{e}-Stark model, Phys. Rev. B {\bf110}, 024207 (2024).
\bibitem{Sun2025}Y.-M. Sun, X.-Y. Wang, and L.-J. Zhai, Non-equilibrium dynamics of localization phase transition in the non-Hermitian disorder-Aubry-Andr\'{e} model, Phys. Rev. A {\bf 112}, 022204 (2025).

\bibitem{Wei2019}B.-B. Wei, Fidelity susceptibility in one-dimensional disordered lattice models, Phys. Rev. A {\bf99}, 042117 (2019).


\bibitem{Bauer1990}J. Bauer, T. M. Chang, and J. L. Skinner, Correlation length and inverse-participation-ratio exponents and multifractal structure for anderson localization, Phys. Rev. B {\bf42}, 8121 (1990).
\bibitem{Fyodorov1992}Y. V. Fyodorov and A. D. Mirlin, Analytical derivation of the scaling law for the inverse participation ratio in quasi-one-dimensional disordered systems, Phys. Rev. Lett. {\bf69}, 1093 (1992).
%

\bibitem{Roati2008}G. Roati, C. D'Errico, L. Fallani, M. Fattori, C. Fort, M. Zaccanti, G. Modugno, M. Modugno and M. Inguscio, Anderson localization of a non-interacting Bose-Einstein condensate, Nature {\bf453}, 895-898 (2008).
\bibitem{Billy2008}J. Billy, V. Josse, Z. Zuo, A. Bernard, B. Hambrecht, P. Lugan, D. Cl\'{e}ment, L. Sanchez-Palencia, P. Bouyer, A. Aspect, Direct observation of Anderson localization of matter waves in a controlled disorder, Nature {\bf453}, 891-894(2008).
%
%
%\bibitem{Wang2021}Y. Wang, C. Cheng, X.-J. Liu, and D. Yu, Many-Body Critical Phase: Extended and Nonthermal, Phys. Rev. Lett. {\bf126}, 080602 (2021).
%\bibitem{Borgnia1}D. S. Borgnia and R.-J. Slager, Localization as a consequence of quasiperiodic bulk-bulk correspondence, Phys. Rev. B {\bf107}, 085111.
%\bibitem{GYSun2022}T. Lv, T.-C. Yi, L. Li, G. Sun and W.-L. You, Quantum criticality and universality in the p-wave-paired Aubry-Andre-Harper model, Phys. Rev. A {\bf105}, 013315 (2022).
%\bibitem{Wang014206}T.Wang, T. Ohtsuki, and R. Shindou, Universality classes of the Anderson transition in the three-dimensional symmetry classes AIII, BDI, C, D, and CI, Phys. Rev. B {\bf104}, 014206 (2021).
%\bibitem{Cestari2011}J. C. C. Cestari, A. Foerster, M. A. Gusm\~{a}o, and M. Continentino, Critical exponents of the disorder driven superfluid-insulator transition in one-dimensional Bose-Einstein condensates, Phys. Rev. A {\bf84}, 055601 (2011).
%
%\bibitem{Billy2008}J. Billy, V. Josse, Z. Zuo, A. Bernard, B. Hambrecht, P. Lugan, D. Cl\'{e}ment, L. Sanchez-Palencia, P. Bouyer, and A. Aspect, Direct observation of Anderson localization of matter waves in a controlled disorder, Nature (London) {\bf453}, 891 (2008).
%\bibitem{Linrj2023}R. -J. Lin, T. Tai, L.-H. Li, C. H. Lee, Topological non-Hermitian skin effect, Front. Phys. {\bf18}, 53605 (2023).
%
%
%
%\bibitem{xueNC2022}Q. Lin, T.-Y Li, L. Xiao, K. -K Wang, W. Yi, P. Xue, Observation of non-Hermitian topological Anderson insulator in quantum dynamics, Nat. Commun. {\bf13}, 3229 (2022).
%\bibitem{Wangxh2024}X. H. Wang and J. Wang, Mpemba effects in nonequilibrium open quantum systems, Phys. Rev. Research {\bf6}, 033330 (2024).
%\bibitem{ChenJJ2024}J. J. Chen and K. Dorfman, Entangled two-photon quantum heat engine: Dissipative nonequilibrium dynamics and correlated statistics, Phys. Rev. Research {\bf6}, 023237 (2024).
%\bibitem{Xue2024}H. X. Gao, K. K. Wang, L. Xiao, M. Nakagawa, N. Matsumoto, D. K. Qu, H. Q. Lin, M. Ueda, and P. Xue, Experimental Observation of the Yang-Lee Quantum Criticality in Open Quantum Systems, Phys. Rev. Lett. {\bf132}, 1776601 (2024).
%\bibitem{Zhang2023131} K. Zhang, C. Fang, and Z. S. Yang, Dynamical Degeneracy Splitting and Directional Invisibility in Non-Hermitian Systems, Phys. Rev. Lett. {\bf131}, 036402 (2023).
%\bibitem{You2024109}L.-T. You, Y.-J. Gao, G.-T. Wang, and G.-P. Zheng, Chiral dynamics of three-mode non-Hermitian systems with a periodical driving, Phys. Rev. A {\bf109}, 062218 (2024).
%\bibitem{Mak2024}J. M\'{a}k, M. J. Bhaseen and A. Pal, Statics and dynamics of non-Hermitian many-body localization, Commun. Phys. {\bf7}, 92 (2024).
%\bibitem{Lizhen2024}Z. Li, L.-W. Wang, X. L. Wang, Z.-K. Lin, G. C. Ma, and J.-H. Jiang, Observation of dynamic non-Hermitian skin effects, Nat. Commun. {\bf15}, 6544 (2024).
%
%\bibitem{Heyl2018}M. Heyl, Dynamical quantum phase transitions: a review, Rep. Prog. Phys. {\bf81}, 054001 (2018).
%\bibitem{Bertini2024}B. Bertini, P. Kos, and T. Prosen, Localized Dynamics in the Floquet Quantum East Model, Phys. Rev. Lett. {\bf132}, 080401 (2024).
%\bibitem{Khan2024}A. Khan, W. Chen, M. Jan, G. Xianlong, Linear-scale simulations of quench dynamics, Comput. Phys. Commun. {\bf132}, 109132 (2024).
%\bibitem{Yin2018}H. H. Yin, S. Chen, X. L. Gao, and P. Wang, Zeros of Loschmidt echo in the presence of Anderson localization, Phys. Rev. A {\bf97}, 033624 (2018).
%\bibitem{Modak2021}R. Modak and D. Rakshit, Many-body dynamical phase transition in a quasiperiodic potential, Phys. Rev. B {\bf103}, 224310 (2021).
%\bibitem{Yang2019}K. Yang, L. Zhou, W. Ma, X. Kong, P. Wang, X. Qin, X. Rong, Y. Wang, F. Shi, J. Gong, and J. Du, Floquet dynamical quantum phase transitions, Phys. Rev. B {\bf100}, 085308 (2019).
%\bibitem{ZhouDu2021}L. W. Zhou and Q. Q. Du, Floquet dynamical quantum phase transitions in periodically quenched systems, J. Phys.: Condens. Matter {\bf33}, 345403 (2021).
%
%\bibitem{Longhi2019}S. Longhi, Topological phase transition in non-Hermitian quasicrystals, Phys. Rev. Lett. {\bf122}, 237601 (2019).
%\bibitem{Luo2021}X. Luo, T. Ohtsuki, and R. Shindou, Universality Classes of the Anderson Transitions Driven by Non-Hermitian Disorder, Phys. Rev. Lett. {\bf126}, 090402 (2021).
%\bibitem{Jing2024}Y. C. Jing, J.-J. Dong, Y.-Y. Zhang, and Z.-X. Hu, Biorthogonal Dynamical Quantum Phase Transitions in Non-Hermitian Systems, Phys. Rev. Lett. {\bf132}, 220402 (2024).
%\bibitem{Yuto2021}Y. Ashida, Z. Gong, and M. Ueda, Non-Hermitian physics, Adv. Phys. {\bf69}, 249 (2021).
%\bibitem{Lv2022}C. W. Lv, R. Zhang, Z. Z. Zhai, and Qi Zhou, Curving the space by non-Hermiticity, Nat. Commun. {\bf13}, 2184 (2022).
%\bibitem{Jiang2023}H. Jiang and C. H. Lee, Dimensional Transmutation from Non-Hermiticity, Phys. Rev. Lett. {\bf131}, 076401 (2023).
%\bibitem{Han2023}P.-R. Han, F. Wu, X.-J. Huang, H.-Z. Wu, C.-L. Zou, W. Yi, M. Z. Zhang, H. K. Li, K. Xu, D. N. Zheng, H. Fan, J. M. Wen, Z.-B. Yang, and S.-B. Zheng, Exceptional Entanglement Phenomena: Non-Hermiticity Meeting Nonclassicality, Phys. Rev. Lett.  {\bf131}, 260201 (2023).
%\bibitem{Sarkar2023}R. Sarkar, A. Bandyopadhyay, and A. Narayan, Non-Hermiticity induced exceptional points and skin effect in the Haldane model on a dice lattice, Phys. Rev. B {\bf107}, 035403 (2023).
%\bibitem{Sunwy2024}W. Y. Sun, L. Luo, Y. Y. Huang, J. B. Peng, D. G. Zhao, Y. W. Yao, F. G. Wu, and X. Zhang, Observation of acoustic hybrid-order topological insulator induced by non-Hermiticity and anisotropy, Phys. Rev. B {\bf109}, 134302 (2024).
%\bibitem{XiaoL2024}L. Xiao, Y. M. Chu, Q. Lin, H. Q. Lin, W. Yi, J. M. Cai, and P. Xue, Non-Hermitian Sensing in the Absence of Exceptional Points, Phys. Rev. Lett. {\bf133}, 180801 (2024).
%\bibitem{Daitx2024}T. X. Dai, Y. T. Ao, J. Mao, Y. Yang, Y. Zheng, C. H. Zhai, Y. D. Li, J. Z. Yuan, B. Tang, Z. H. Li, J. Luo, W. W. Wang, X. Y. Hu, Q. H. Gong and J. W. Wang, Non-Hermitian topological phase transitions controlled by nonlinearity, Nat. Phys. {\bf20}, 101 (2024).
%\bibitem{Zhang2109431}X. Zhang, T. Zhang, M.-H. Lu, and Y.-F. Chen, A review on non-Hermitian skin effect, Adv. Phys.: X {\bf7}, 2109431 (2022).
%\bibitem{Liu2024}G.-G. Liu, S. Mandal, P. H. Zhou, X. Xi, R. Banerjee, Y.-H. Hu, M. G. Wei, M. R. Wang, Q. Wang, Z. Gao, H. S. Chen, Y. H. Yang, Y. D. Chong, and B. Zhang, Localization of Chiral Edge States by the Non-Hermitian Skin Effect, Phys. Rev. Lett. {\bf132}, 113802 (2024).
%\bibitem{Gliozzi2024}J. Gliozzi, G. D. Tomasi, and T. L. Hughes, Many-Body Non-Hermitian Skin Effect for Multipoles, Phys. Rev. Lett. {\bf133}, 136503 (2024).
%\bibitem{YoshidaT2024}T. Yoshida, S.-B. Zhang, T. Neupert, and N. Kawakami, Non-Hermitian Mott Skin Effect, Phys. Rev. Lett. {\bf133}, 076502 (2024).
%\bibitem{Wittrock2024}S. Wittrock, S. Perna, R. Lebrun, K. Ho, R. Dutra, R. Ferreira, P. Bortolotti, C. Serpico and V. Cros, Non-hermiticity in spintronics: oscillation death in coupled spintronic nano-oscillators through emerging exceptional points, Nat. Commun. {\bf15}, 971 (2024).
%\bibitem{Wang024202}C. Wang and X. R. Wang, Anderson localization transitions in disordered non-hermitian systems with exceptional points, Phys. Rev. B {\bf107}, 024202 (2023).
%\bibitem{Martinez2018}V. M. Martinez Alvarez, J. E. Barrios Vargas, L. E. F. Foa Torres, Non-Hermitian robust edge states in one dimension: Anomalous localization and eigenspace condensation at exceptional points, Phys. Rev. B {\bf97}, 121401(R) (2018).
%\bibitem{Aodong706}A. Li, H. Wei, M. Cotrufo, W. Chen, S. Mann, X. Ni, B. Xu, J. Chen, J. Wang, S. Fan, C.-W. Qiu, A. Al, and L. Chen, Exceptional points and non-Hermitian photonics at the nanoscale, Nat. Nanotechnol. {\bf18}, 706 (2023).
%\bibitem{Wingenbach2024}J. Wingenbach, S. Schumacher, and X. K. Ma, Manipulating spectral topology and exceptional points by nonlinearity in non-Hermitian polariton systems, Phys. Rev. Research {\bf6}, 013148 (2024).
%
%\bibitem{Kochergin2024}D. Kochergin, V. Tiselko, and A. Onuchin, Localization transition in non-Hermitian systems depending on reciprocity and hopping asymmetry, Phys. Rev. E  {\bf109},, 044315 (2024).
%
%\bibitem{Hatano1996}N. Hatano and D. R. Nelson, Localization transitions in non-Hermitian quantum mechanics, Phys. Rev. Lett. {\bf77}, 570 (1996).
%\bibitem{Hatano1998}N. Hatano and D. R. Nelson, Non-Hermitian delocalization and eigenfunctions, Phys. Rev. B {\bf58}, 8384 (1998).
%\bibitem{Luo0153523}X.-W. Luo and C. Zhang, Photonic topological insulators induced by non-Hermitian disorders in a coupled-cavity array, Appl. Phys. Lett. {\bf123}, 081111 (2023).
%\bibitem{Chen144208}W. Chen, S. Cheng, J. Lin, R. Asgari, and X. Gao, Breakdown of the correspondence between the real-complex and delocalization-localization transitions in non-Hermitian quasicrystals, Phys. Rev. B {\bf106}, 144208 (2022).
%\bibitem{Wang024514}Z.-H. Wang, F. Xu, L. Li, D.-H. Xu, and B. Wang, Topological superconductors and exact mobility edges in non-Hermitian quasicrystals, Phys. Rev. B {\bf105}, 024514 (2022).
%\bibitem{Jazaeri2001}A. Jazaeri and I. I. Satija, Localization transition in incommensurate non-Hermitian systems, Phys. Rev. E {\bf63}, 036222 (2001).
%\bibitem{PWang2019}P. Wang, L. Jin, and Z. Song, Non-Hermitian phase transition and eigenstate localization induced by asymmetric coupling, Phys. Rev. A {\bf99}, 062112 (2019).
%
%\bibitem{Chaohua123048}C. Wu, J. Fan, G. Chen, and S. Jia, Non-Hermiticity-induced reentrant localization in a quasiperiodic lattice, New J. Phys. {\bf23}, 123048 (2021).
%\bibitem{YCWang2023}Y.-C. Wang, K. Suthar, H. H. Jen, Y.-T. Hsu, and J.-S. You, Non-Hermitian skin effects on thermal and many-body localized phases, Phys. Rev. B {\bf107}, L220205 (2023).
%\bibitem{KSuthar2022}K. Suthar, Y.-C. Wang, Y.-P. Huang, H. H. Jen, and J.-S. You, Non-Hermitian many-body localization with open boundaries, Phys. Rev. B {\bf106}, 064208 (2022).
%\bibitem{Jiang2024}X.-P. Jiang, W. Zeng, Y. Hu, and P. Liu, Exact non-Hermitian mobility edges and robust flat bands in two-dimensional Lieb lattices with imaginary quasiperiodic potentials, New J. Phys. {\bf26}, 083020 (2024).
%\bibitem{Hamazaki2019} R. Hamazaki, K. Kawabata, and M. Ueda, Non-Hermitian many-body localization, Phys. Rev. Lett. {\bf123}, 090603 (2019).
%
%
%
%\bibitem{Zhong1}F. Zhong, Probing criticality with linearly varying external fields: Renormalization group theory of nonequilibrium critical dynamics under driving, Phys. Rev. E {\bf73}, 047102 (2006).
%\bibitem{Zhong2}S. Gong, F. Zhong, X. Huang, and S. Fan, Finite-time scaling via linear driving, New J. Phys. {\bf12}, 043036 (2010).
%\bibitem{Zhong3}Y. Huang, S. Yin, B. Feng, and F. Zhong, Kibble-Zurek mechanism and finite-time scaling, Phys. Rev. B {\bf90}, 134108 (2014).
%
%
%\bibitem{Tomasi2023}G. D. Tomasi and I. M. Khaymovich, Non-Hermiticity induces localization: Good and bad resonances in power-law random banded matrices, Phys. Rev. B {\bf108}, L180202 (2023).
%\bibitem{Longhi2023}S. Longhi, Non-Hermitian control of localization in mosaic photonic lattices, Appl. Phys. Lett. {\bf123}, 161102 (2023).


\end{thebibliography}
\end{document}